\documentclass[twocolumn,aps,groupedaddress]{revtex4}

\usepackage[colorlinks=true,citecolor=blue,linkcolor=magenta]{hyperref}
\usepackage{graphicx}
\usepackage{dcolumn}
\usepackage{bm}
\usepackage{amsmath,amssymb,amsfonts}
\usepackage{color}
\usepackage{ulem}

\newcommand{\be}{\begin{align}}
\newcommand{\ee}{\end{align}}
\def \beq{\begin{equation}}
\def \eeq{\end{equation}}
\def \ba{\begin{array}}
\def \ea{\end{array}}
\def \bea{\begin{eqnarray}}
\def \eea{\end{eqnarray}}
\def \nn{\nonumber}
\def \ba{\begin{align*}}
\def \ea{\end{align*}}

\begin{document}

\title{Designing Flat Band by Strain}

\author{Zhen Bi}
\email{zbi@mit.edu}
\author{Noah F. Q. Yuan}
\author{Liang Fu}
\affiliation{Department of Physics, Massachusetts Institute of Technology, Cambridge, MA 02139 USA}
\begin{abstract}
We study the effects of heterostrain on moir\'e bands in twisted bilayer graphene and bilayer transition metal dichalcogenide (TMD) systems. For bilayer graphene with twist angle near $1^\circ$, we show that heterostrain significantly increases the energy separation between conduction and valence bands as well as the Dirac velocity at charge neutrality, which resolves several puzzles in scanning tunneling spectroscopy and quantum oscillation experiments at once. For bilayer TMD, we show that applying small heterostrain generally leads to flat moir\'e bands that are highly tunable.
\end{abstract}
\maketitle
\section{Introduction}

Recent experimental discoveries of correlated insulator and superconducting states in two-dimensional moir\'e materials including twisted bilayer graphene (TBG)\cite{TBGexp1,TBGexp2,TBGexp3} and graphene-boron nitride heterostructures \cite{TLGexp1} have stimulated tremendous interest in engineering flat or narrow bands to realize correlated electron phenomena. For this purpose, moir\'e superlattices provide a unique and highly tunable material platform. With a vast variety of $2d$ materials and heterostructures available \cite{2dreview}, moir\'e superlattice systems may offer unprecedented advantages for studying many-body physics  and realizing exotic states of matter. 

Moir\'e patterns appear ubiquitously in layered $2d$ materials with a slight mismatch in the lattice orientation and/or lattice constant of the layers. Aside from twist angles between layers, heterostrain---which refers to relative strains between layers---provides an alternative way to create and modify moir\'e patterns\cite{heterostrainPRL}. Using strain to tune moir\'e bands may have practical advantages than using twist angle. Strain can be controlled \textit{in situ} via piezoelectric substrate. The possibility of tuning band structure and achieving partially flat band with strain has recently been studied in graphene (see for example \cite{straineng,straineng2}) and surface states of topological crystalline insulator \cite{TangFu}.

In this work, we systematically study the effect of heterostrain on moir\'e band structures for homobilayer systems including bilayer graphene and transition metal dichalcogenides (TMD). Our motivation is twofold. First, (unintentional) heterostrain is ubiquitous in TBG samples \cite{TBGstm}, likely due to the interaction with the substrate. However, the effect of heterostrain on the flat bands in magic-angle TBG is not well understood. We find a small amount of strain dramatically changes the low-energy band structure, which resolves several puzzles in transport and spectroscopy experiments all at once.  Second, engineering a {\it tunable flat band} system by strain provides unprecedented opportunities for future studies. To that end, we focus on bilayer TMDs and show the conditions for flat moir\'e bands induced by heterostrain without any twist.

This work consists of two parts.
In the first part, we show that heterostrain in TBG significantly increases the energy separation between moir\'e conduction and valence bands, for instance, to $\sim30$meV with $0.5\%$ uniaxial heterostrain as commonly observed in STM experiments \cite{TBGstm}. The energy separation of the van Hove singularities saturates over a range of twist angles $\theta$ around $1^\circ$, rather than being extremely sensitive to small deviation from the magic angle in the unstrained case. These findings explain the unexpectedly large separation of van Hove singularities observed in the recent STM experiments on TBG with $\theta \sim 1^\circ$ \cite{TBGstm}. 

Moreover, in the presence of heterostrain, the conduction and valence bands of the same valley remain connected by two Dirac points near charge neutrality. However, due to the lowered symmetry, these two Dirac points are away from mini-Brillouin zone corners and are no longer degenerate in energy. This effect may explain the 4 (instead of 8) fold Landau level degeneracy\footnote{The possibility of the lifted degeneracy by heterostrain is brought to our attention by Matthew Yankowitz and Cory Dean.} observed in the transport experiments\cite{TBGexp1,TBGexp2,TBGexp3}. We also show that heterostrain sets a lower bound for the Dirac velocity, preventing it from vanishing at the magic angle. The Dirac velocity is anisotropic and on the order of $0.14v_F$ for a small heterostrain of $0.3\%$-$0.6\%$ 
($v_F$ being the bare Dirac velocity of monolayer graphene). This value is comparable to the one inferred from quantum oscillation and capacitance measurements\cite{TBGexp1}.

In the second part, we study bilayer TMDs with volume preserving heterostrain. In contrast to the twisted bilayer graphene, we find that nearly flat moir\'e bands are generically present near the top of valence band in heterostrained bilayer TMD without fine tuning and without the need of twist. In addition, the moir\'e band gap and band structure are highly tunable by strain, pressure and displacement field, thus providing an ideal platform for correlation-driven phenomena.

\section{2-dimensional strain and general continuum models}

In this section, we consider the general geometrical description of $2$-dimensional strain. Mathematically, the coordinate transformation in $2d$ can be written as
\beq
\bold{r}'=(\mathbb{I}+\mathcal{E})\bold{r}+\bold{d}_0,
\eeq
where $\bold{d}_0$ is a $2$-dimensional vector that parametrizes the displacement and $\mathcal{E}$ is an arbitrary $2$-dimensional matrix that contains the strain and rotation. In the small deformation limit,  $\mathcal{E}$ can be written as
\beq
\mathcal{E}\cong
\left(\begin{matrix}
\epsilon_{xx} & \epsilon_{xy}-\theta \\
\epsilon_{xy}+\theta & \epsilon_{yy}
\end{matrix}
\right)=\mathcal{S}(\epsilon)+\mathcal{T}(\theta),
\eeq
where the symmetric part, labeled by $\mathcal{S}(\epsilon)$, represents the strain, and the anti-symmetric part, labeled by $\mathcal{T}(\theta)$, represents the rotation.  We emphasis that strain for $2d$ materials, different from the rotation angle which is a single parameter, is characterized by a 2 by 2 symmetric matrix, which has 3 free parameters. Therefore, it provides more possibilities to engineer the superlattice structures for layered $2d$ materials.

Let us review the effects of geometric deformation on the properties of monolayer graphene. For free monolayer graphene, we define $\bold{A}_1=a(1,0)$ and $\bold{A}_2=a(1/2,\sqrt{3}/2)$ to be the primitive lattice vectors. Correspondingly, the reciprocal lattice vectors are $\bold{G}_1=\frac{2\pi}{a}(1,-1/\sqrt{3})$ and $\bold{G}_2=\frac{2\pi}{a}(0,2/\sqrt{3})$. $\bold{K}_\pm=\mp(2\bold{G}_1+\bold{G}_2)/3$ are referred as two valley points. The low energy description of monolayer graphene contains two massless Dirac fermions at $\bold{K}_+$ and $\bold{K}_-$ points with spin degeneracy. Geometrically, a deformation $\mathcal{E}$ changes the shape of the unit cell as well as the Brillouin zone. Mathematically the rescaled primitive and reciprocal lattice vectors are
\beq
\bold{A}'_i=(\mathbb{I}+\mathcal{E})\bold{A}_i, \ \ \bold{G}'_i\cong (\mathbb{I}-\mathcal{E}^T)\bold{G}_i.
\eeq
A generic $\mathcal{E}$, including both nonzero strain $\mathcal{S}$ and rotation $\mathcal{T}$, breaks almost all the point group symmetry of the lattice except the inversion $C_{2z}$. In addition to the geometric effect, the strain $\mathcal{S}$ adjusts the distances between atoms, which leads to differences in hopping matrix elements for nearest carbon atoms along different directions. As a result, the locations of the low energy Dirac fermions are shifted away from the rescaled valley points $\bold{K}'_\pm\cong (\mathbb{I}-\mathcal{E}^T)\bold{K}_\pm$. Within a simple two center approximation $t(r)\sim t_0e^{\beta(r/a_0-1)}$, the shift is proportional to the strain in the small strain limit and can be described by an effective gauge connection for the low energy Dirac fermions\cite{KoshinoStrain}:
\beq
\bold{A}=\frac{\sqrt{3}}{2a}\beta(\epsilon_{xx}-\epsilon_{yy},-2\epsilon_{xy})
\label{eq:shift}
\eeq
The two Dirac fermions carry opposite charges under this fictitious gauge field. Therefore, the positions of the two Dirac fermions in momentum space are given by
\beq
\bold{D}_\xi=(\mathbb{I}-\mathcal{E}^T)\bold{K}_\xi-\xi \bold{A},
\label{eq:dirac}
\eeq
where $\xi=\pm$ labels the two valleys. The hopping modulus factor $\beta$ is a dimensionless parameter determined by the intrinsic properties of the material. The approximate value of $\beta$ is estimated by first principle calculation for graphene, $\beta_g\cong 3.14$\cite{KoshinoStrain}.

The effects of geometric deformation for monolayer TMDs are similar as for monolayer graphene. For unstrained monolayer TMD, the low energy theory can be modeled by two \textit{massive} Dirac fermions\cite{XDTMD1,YWTMD1} at $\bold{K}_\pm$ points. As a result of the large spin-orbital coupling, the valence bands near two valleys carry opposite spins. Generic deformations reshape the unit cell and the Brillouin zone and break all the point group symmetries. It also shifts the Dirac fermions away from the rescaled valley points $\bold{K}'_\pm$. The shift is again described by Eq. \ref{eq:shift} and \ref{eq:dirac}. First principle calculations suggest for WSe$_2$, the hopping modulus factor is $\beta_{WSe_2}\cong 2.30$\cite{Tstrain}.

In the rest of the paper, we consider homobilayers systems starting with \textit{AA stacking} and then apply small twist and heterostrain. The coordinate transformations of the two layers can be described by two deformation matrices $\mathcal{E}_1$ and $\mathcal{E}_2$. The rescaled reciprocal lattice vectors for the two layers are
\beq
\bold{G}'_i\cong(\mathbb{I}-\mathcal{E}_1^T)\bold{G}_i, \ \ \bold{G}''_i\cong(\mathbb{I}-\mathcal{E}_2^T)\bold{G}_i.
\eeq
Such geometrical deformation generates a moir\'e superlattice whose reciprocal lattice vectors $\bold{g}_i$'s and primitive lattice vectors $\bold{a}_i$'s are given by
\beq
\bold{g}_i\cong\bold{G}'_i-\bold{G}''_i=\mathcal{E}^T\bold{G}_i,\ \ \bold{a}_i\cong\mathcal{E}^{-1}\bold{A}_i.
\eeq
respectively, where $\mathcal{E}=\mathcal{E}_2-\mathcal{E}_1$ is the relative deformation matrix. Physical properties only depend on the relative deformation $\mathcal{E}$ in the small twist and strain limit. For all the calculations in this paper, we assume that $\mathcal{E}_2=-\mathcal{E}_1=\frac{1}{2}\mathcal{E}$, namely the two layers are rotated and strained oppositely with the same magnitude.

In the limit where the moir\'e superlattice constant is much larger than the atomic scale, the low energy electronic structure can be effectively captured by the continuum model\cite{McDCM,NetoCM1,NetoCM2}. The spirit of the continuum model is the same as the nearly free electron approximation for band structures in solid state physics. Essentially, one should take the low energy bare dispersions of the $2d$ materials and perturb them by the periodic moir\'e superlattice. A general continuum model for bilayer systems can be schematically written as\cite{McDCM,FuTBG1,FuTBG2}
\beq
H=\left(\begin{matrix}
h_1(\bold{k})+V_1(\bold{r})& T(\bold{r}) \\
T^\dagger(\bold{r})& h_2(\bold{k})+V_2(\bold{r})
\end{matrix}
\right),
\label{eq:H}
\eeq
where $h_1(\bold{k})$ and $h_2(\bold{k})$ are the bare dispersions of layer 1 and 2 respectively. The $T(\bold{r})$ describes the spatial dependent interlayer tunneling and the $V_1(\bold{r})$ and $V_2(\bold{r})$ describe the intralayer potential induced by the moir\'e superlattice. The $T(\bold{r})$ and $V(\bold{r})$ vary between different materials and are also dependent on extrinsic conditions such as pressure\cite{TBGexp3}. In the following sections, we will focus our discussions on two cases $\mathit{1}^\circ$ twisted bilayer graphene with uniaxial heterostrain and $\mathit{2}^\circ$ bilayer WSe$_2$ with volume preserving heterostrain.

\section{Twisted bilayer graphene with uniaxial heterostrain}

\subsection{The continuum model}

In bilayer graphene, the bare low energy dispersion for each layer contains two massless Dirac fermions, namely
\beq
h_l(\bold{k})=\sum_{\xi=\pm}-\hbar v_F/a[(\mathbb{I}+\mathcal{E}_l^T)(\bold{k}-\bold{D}_{l,\xi})]\cdot (\xi \sigma^x,\sigma^y),
\eeq
where $l=1,2$ labels the two layers, $\xi=\pm$ labels the two valleys. The $\sigma$ matrices act on the pseudospin or sublattice degrees of freedom. Each valley also has two-fold spin degeneracy. $\bold{D}_{l,\xi}$ denotes the location of the Dirac fermion as given in Eq. \ref{eq:dirac}. The fermi velocity in monolayer graphene is estimated $v_F\cong 10^6$ m/s\cite{grapheneRMP}, which gives the kinetic energy scale $\hbar v_F/a\cong 2.68$eV.

In this section, we restrict ourselves to twisted bilayer graphene with uniaxial heterostrain. The uniaxial heterostrain refers to a class of strain where the bilayer system is relatively stressed along one direction and unstressed on the perpendicular direction. Scanning tunneling microscope (STM) experiments have indicated such type of heterostrain in twisted bilayer graphene samples\cite{TBGstm, TBGstm2, TBGstm3}. We anticipate that many generic features of the uniaxial strain should also apply for more general form of strain. Geometrically, uniaxial strain can be described by two parameters, namely the strain magnitude $\epsilon$ and the strain direction $\varphi$. The strain tensor can be written as the following\cite{TBGstm}
\bea
\mathcal{S}_{ua}&=&R(\varphi)^{-1}
\left(\begin{matrix}
-\epsilon& 0 \\
0 & \nu\epsilon
\end{matrix}
\right)R(\varphi) \\ \nonumber
&=&
\epsilon\left(\begin{matrix}
-\cos(\varphi)^2 +\nu \sin(\varphi)^2& (1+\nu)\cos(\varphi)\sin(\varphi) \\
 (1+\nu)\cos(\varphi)\sin(\varphi)& -\sin(\varphi)^2 +\nu \cos(\varphi)^2
\end{matrix}
\right),
\eea
where $\nu=0.16$ is the Poisson ratio for graphene. The combination of twist and strain is described by the relative deformation matrics $\mathcal{E}=\mathcal{T}(\theta)+\mathcal{S}_{ua}(\epsilon,\varphi)$.

For graphene system, $V_l(\bold{r})$ is parametrically smaller than the interlayer tunneling, therefore we set it to zero. The interlayer tunneling has the following form (following the convention in, for example, \cite{FuTBG2}),
\bea
\nonumber
T(\bold{r})=&\left(\begin{smallmatrix}
u & u' \\
u' & u''
\end{smallmatrix}
\right)+\left(\begin{smallmatrix}
u & u' \omega^{-\xi} \\
u' \omega^{\xi} & u''
\end{smallmatrix}
\right)e^{i\xi \bold{g}_1\cdot \bold{r}}\\ 
&+\left(\begin{smallmatrix}
u & u' \omega^{\xi} \\
u' \omega^{-\xi} & u''
\end{smallmatrix}
\right)e^{i\xi (\bold{g}_1+\bold{g}_2)\cdot \bold{r}},
\label{eq:tunneling}
\eea
where $\omega=e^{i2\pi/3}$.  The effect of lattice corrugation in TBG on moir\'e band structure   can be included in the continuum model by choosing renormalized tunneling amplitudes  $u,u'$ and $u''$ \cite{FuTBG2}, which depend on  heterostrain and twist angle. For simplicity, in our calculation we assume the two graphene layers are unrelaxed and use the tunneling parameters $u=u'=u''=110$meV\cite{McDCM}. With this set of parameters, the first magic angle without strain locates around $\theta_c\cong 0.95^\circ$.

\begin{figure}
\begin{center}
\includegraphics[width=0.45\textwidth]{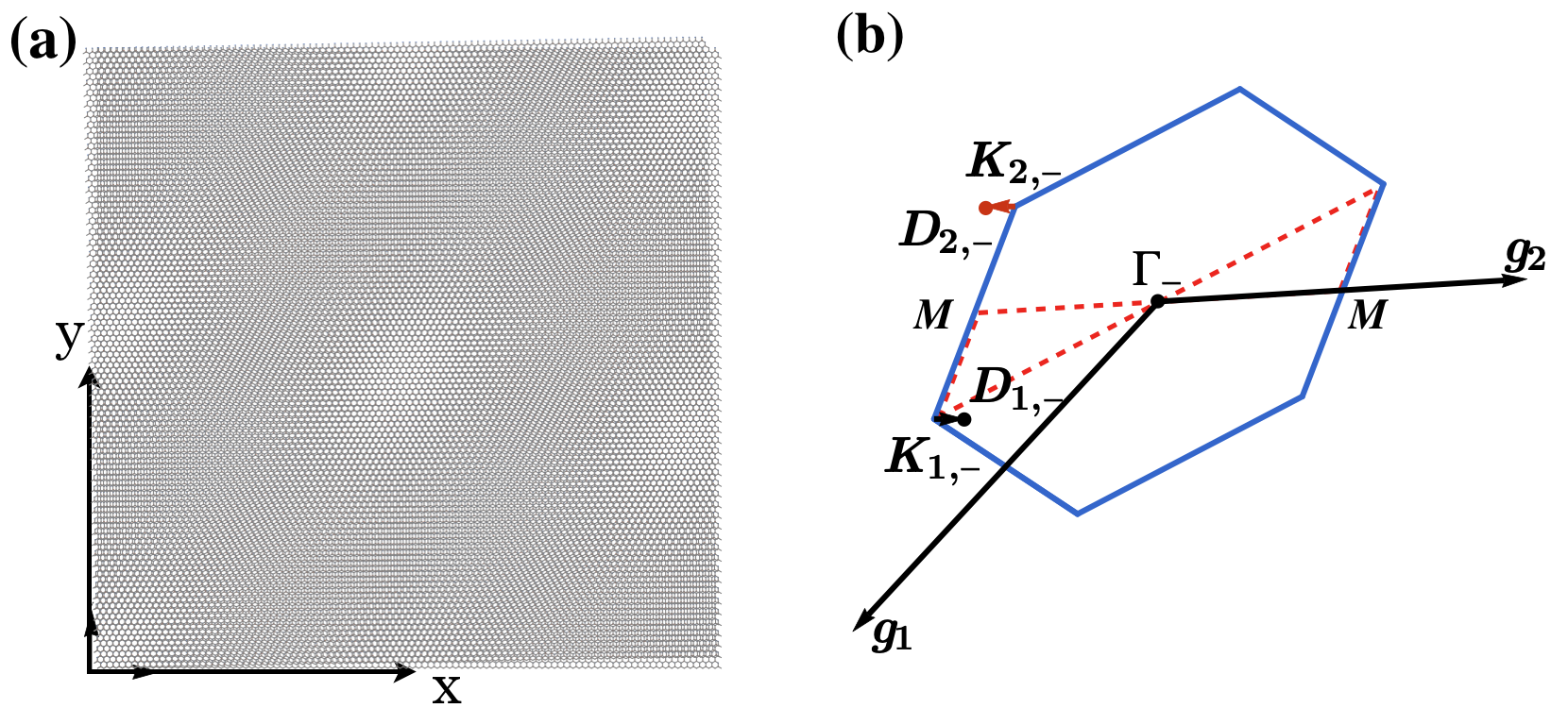}
\end{center}
\caption{(a)The moir\'e pattern of bilayer grahenen with twist angle $\theta=1.05^\circ$ and uniaxial strain $\epsilon=0.7\%$, $\varphi=0^\circ$. Notice that the moir\'e superlattice is not a regular triangular lattice. The moir\'e dots are elliptical which is commonly observed in STM experiments \cite{TBGstm}. (b) The moir\'e Brillouin zone with the same geometrical parameters. In the plot, for clearity, we only label the rescaled $\bold{K}_-$ points from the two layers. Correspondingly, the red and black dots are the position of the shifted Dirac points from $\bold{K}_-$ valley. The other valley can be obtained by time reversal operation. The red dashed line is the momentum path for the band structure plots.}
\label{fig:MP}
\end{figure}

An apparent geometric effect of the combination of twist and heterostrain is that the moir\'e superlattice is no longer a regular triangular lattice . An example of moir\'e pattern generated with $\theta=1.05^\circ$, $\epsilon=0.7\%$ and $\varphi=0^\circ$ is shown in Fig. \ref{fig:MP}(a). The uniaxial strain makes the moir\'e dots elliptical, which is visible in local measurements such as STM\cite{TBGstm}. As a consequence, the Brillouin zone of the moir\'e superlattice is a distorted hexagonal as shown in Fig. \ref{fig:MP}(b). In order to compare results for different twists and strains, we will always stretch the irregular hexagonal Brillouin zone to be regular.

\begin{figure}
\begin{center}
\includegraphics[width=0.45\textwidth]{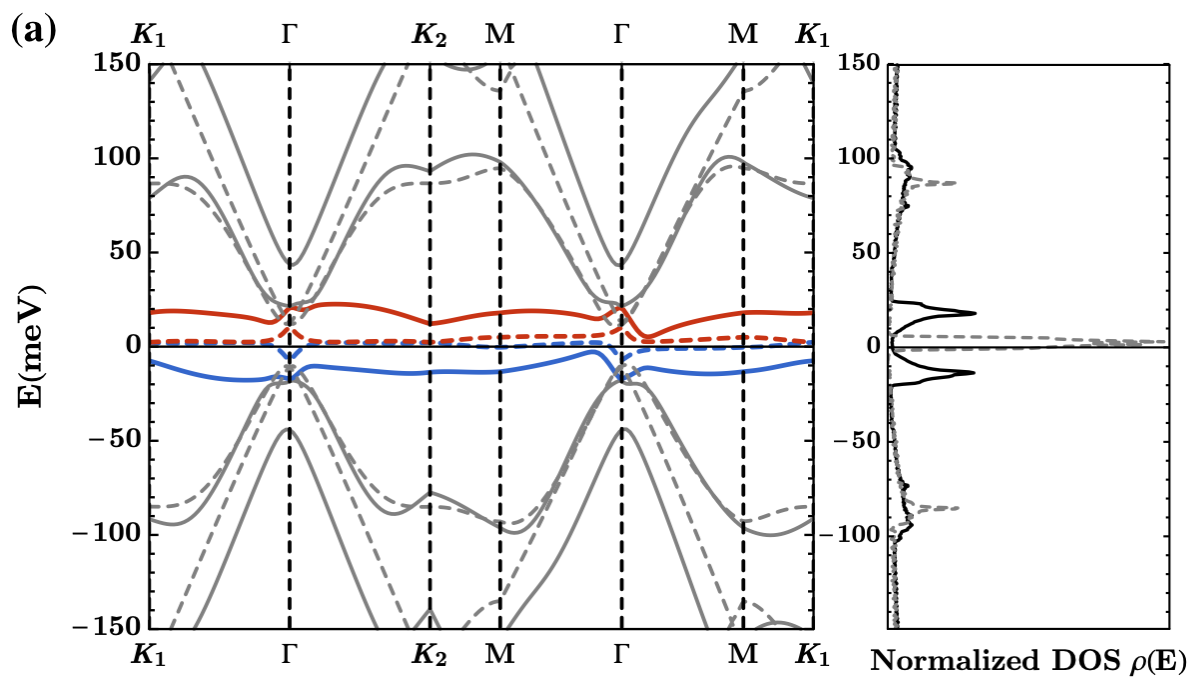}
\includegraphics[width=0.47\textwidth]{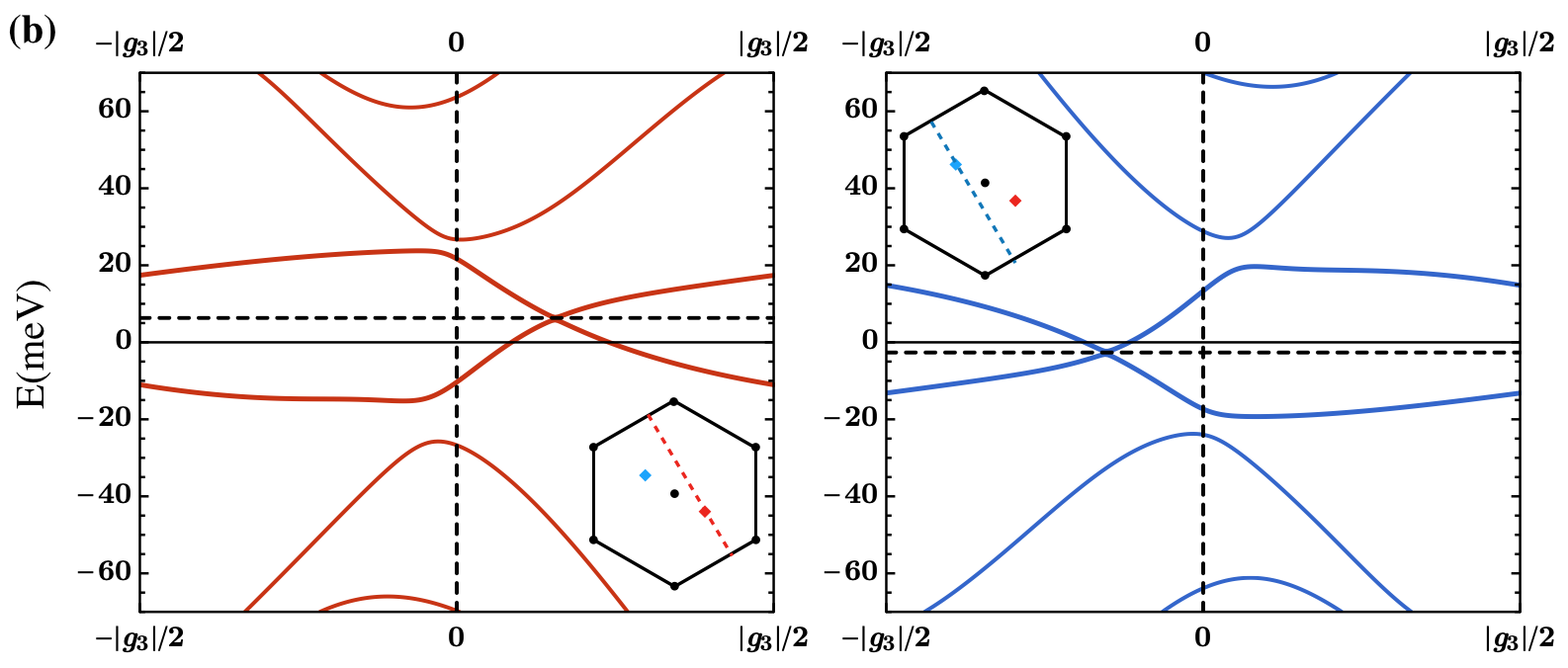}
\end{center}
\caption{In (a), we show the band structure (left) and the density of states (right) for the $\bold{K}_-$ valley of twisted bilayer graphene at twist angle $\theta\cong 1.05^\circ$ with (solid line) and without (dashed line) uniaxial heterostrain $\epsilon=0.6\%$, $\varphi=30^\circ$. The heterostrain increases the separation between the valence and conduction bands, while keeps them rather flat in most area of the Brillouin zone. In (b), we show that the two bands are still connected to each other through two Dirac points. With heterostrain, the positions of the two Dirac crossings are no longer at the corner of the moir\'e Brillouin zone, as shown in the inserts of Fig (b). The hexagons represent the stretched Brillouin zone and the blue/red dots are the positions of two Dirac crossings. We plot the band dispersions along the red/blue dashed lines in the Brillouin zone. The conduction and valence bands are rather flat in most part of the momentum space except near the Dirac crossings. The energy shift between the two Dirac points is $\sim 9$meV. As seen from (b), the Dirac velocity is anisotropic and reaching $0.15v_F$, which is much larger than the unstrained case (about 0.018$v_F$ with the current parameters). This velocity is comparable to the one observed in transport and capacitance experiments\cite{TBGexp1}. Since the Dirac points are shifted in energy, a finite electron/hole fermi surfaces appear at charge neutrality. Due to the enhanced Dirac velocity, the size of the fermi pocket is found to be much smaller than the size of the Brillouin zone. The estimated electron/hole density (including the valley and spin degeneracy) is $\sim3\times10^{10}$ cm$^{-2}$ for current parameters.
}
\label{fig:bands}
\end{figure}

A typical band structure of twisted bilayer graphene near magic angle with small uniaxial heterostrain is shown in Fig. \ref{fig:bands}. Three generic features are worth noticing in these plots. First, the energy separation between conduction and valence bands, as seen from the separation of van Hove singularities, is significantly enlarged compared to the case with no strain. Interestingly, these two bands remain rather flat for most parts of the Brillouin zone. This finding agrees with the large splitting of van Hove singularities of the conduction and valence bands observed in the scanning tunneling spectroscopy (STS) \cite{TBGstm}. We will study in detail of the evolution of the bandwidth with respect to the geometrical parameters $\theta$, $\epsilon$ and $\varphi$.

Second, the conduction and valence bands remain connected by two Dirac crossings in each valley, which is due to the $C_{2z}\mathcal{T}$ symmetry. However, these two Dirac crossings are now located at generic points away from Brillouin zone corners because of the lack of three-fold rotational symmetries in the presence of strain. 
In addition, the energies of the two Dirac points within a valley are shifted by the heterostrain and are no longer degenerate. This may explain the observed 4-fold (instead of 8-fold) Landau level degeneracy in the experiments\cite{TBGexp1,TBGexp2,TBGexp3} near charge neutrality. We note that 
the energy offset of the two Dirac points 
leads to a finite density of electrons and holes at charge neutrality. Nonetheless, the residual electron/hole density is found to be rather small: around $\sim3\times 10^{10}$cm$^{-2}$ (including valley and spin degeneracy)  for $0.6\%$ strain.
We shall also investigate the dependence of the Dirac fermion energy splitting on the system parameters.

Finally, compared to the unstrained case, the Dirac velocity is found to be greatly enhanced---reaching $0.14v_F$ for a small strain of $0.6\%$, and is anistropic due to the lowered symmetry.
Moreover, we find that heterostrain sets a lower bound for the Dirac velocity, preventing it from vanishing around magic angle. The enhanced Dirac velocity found here is comparable to the one inferred from quantum oscillation and capacitance experiments \cite{TBGexp1}.


\subsection{The bandwidth and the Dirac point shift}

Now we study $\mathit{1}^\circ$ the total bandwidth of the conduction and valence bands and $\mathit{2}^\circ$ the energy shift of the two Dirac crossings within a valley as a function of the parameters $\theta$, $\epsilon$ and $\varphi$. We are interested in the regime of twist angle $\theta$ around $1^\circ$. Experimental data indicate that the uniaxial heterostrain in twisted bilayer graphene samples can vary from $0.1\%$ to $0.7\%$\cite{TBGstm}. In addition, it is easy to convince ourselves that the system is periodic for $\varphi\rightarrow \varphi+\pi/3$ because of the symmetry of the unstrained system. Therefore, we can restrict our studies to $\varphi\in[0,\pi/3)$.

\begin{figure}
\begin{center}
\includegraphics[width=0.40\textwidth]{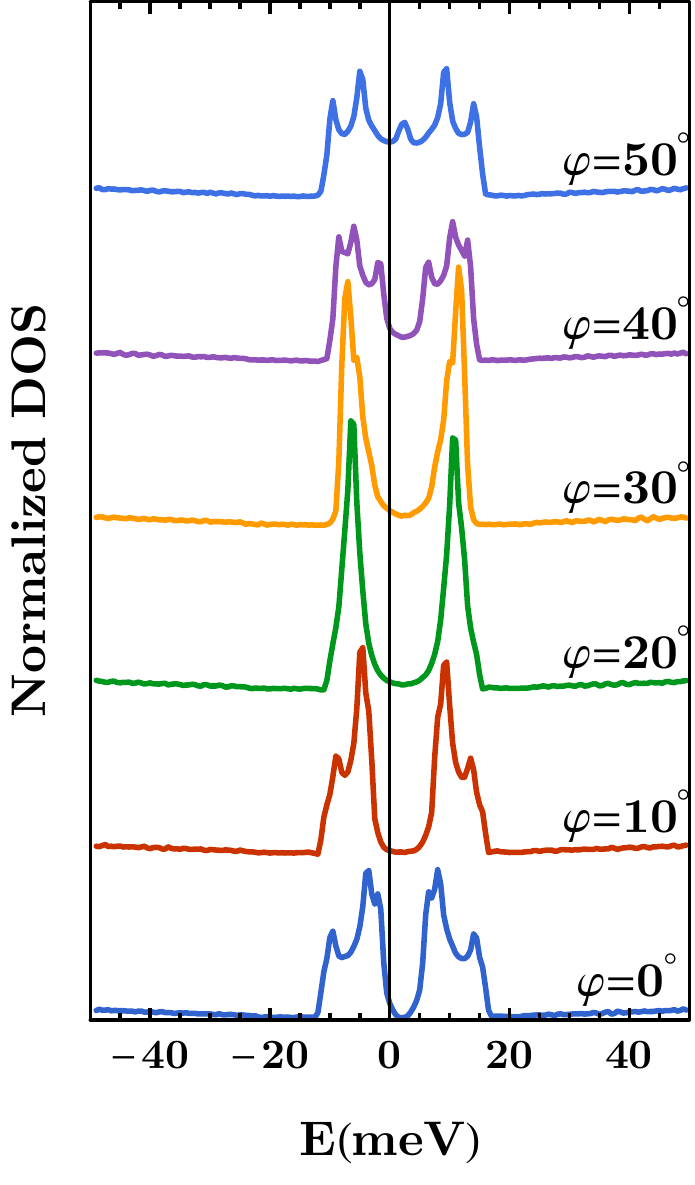}
\end{center}
\caption{Normalized density of state plot with fixed $\theta=1.05^\circ$, $\epsilon=0.3\%$ and varying $\varphi$. (The curves are relatively shifted to make the plot clear.) We observe many subtle transitions in the structure of the van Hove singularities as the direction of the strain is changes. However, the bandwidth of the middle two bands stays approximately constant. (The density of states has only one prominent peak in each band for $\varphi$ near $20^\circ\sim30^\circ$.) }
\label{fig:DOS}
\end{figure}

Firstly, let us fix $\theta=1.05^\circ$ and $\epsilon=0.3\%$ and plot the density of states the conduction and valence bands for several values of $\varphi$ as shown in Fig. \ref{fig:DOS}. We observe many subtle changes in the structure of van Hove singularities within each bands as the direction of the strain is varied. For instance, $\varphi=0^\circ$ has three density of state peaks in each band. However, for $\varphi=20^\circ \sim 30^\circ$, there is only one prominent peak in each band, which is consistent with the experimental results\cite{TBGstm}. Nonetheless, the total bandwidth stays approximately \textit{constant}.

\begin{figure}
\begin{center}
\includegraphics[width=0.38\textwidth]{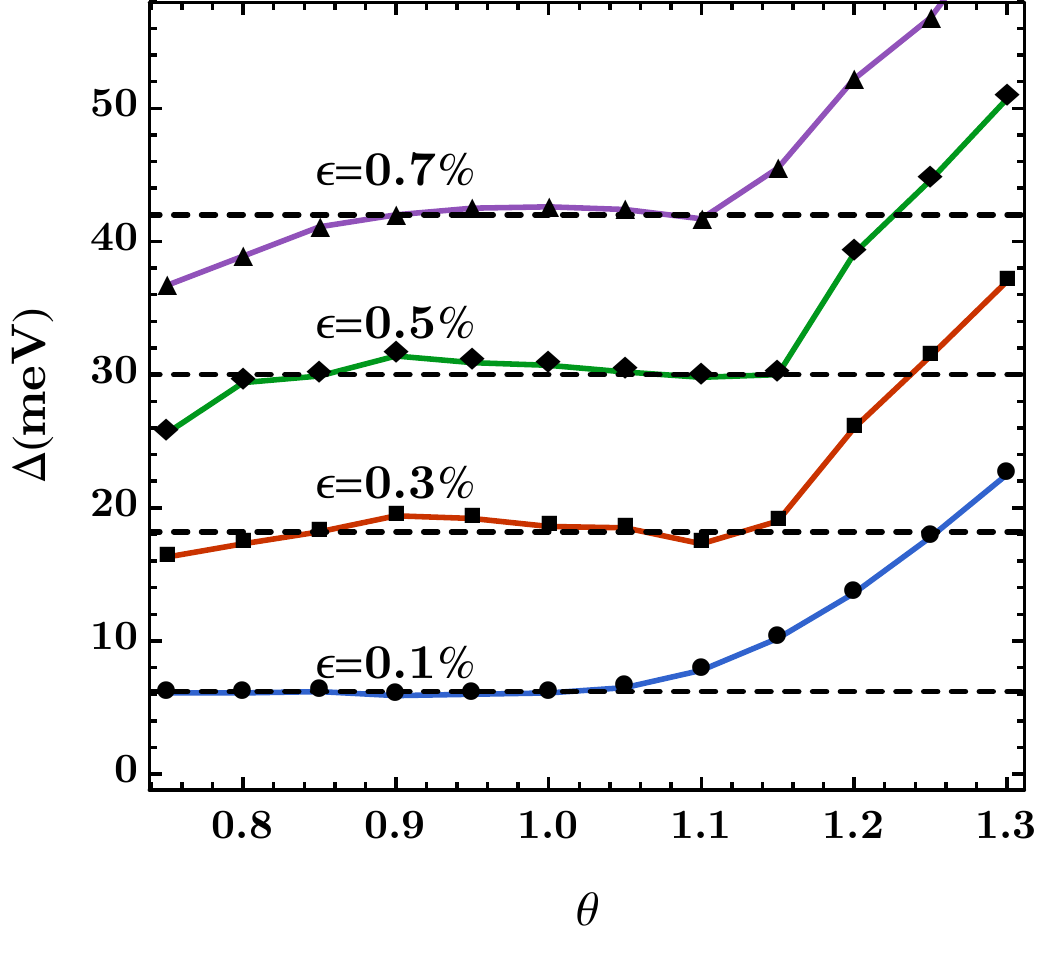}
\end{center}
\caption{The splitting of the van Hove singularities, $\Delta$, as a function of the twist angle $\theta$. We fixed the strain direction $\varphi=25^\circ$ and different curves represent different strain magnitudes. The splitting of the van Hove singularities stay approximately constant for nearly magic twist angles.}
\label{fig:BW2}
\end{figure}

\begin{figure}
\begin{center}
\includegraphics[width=0.4\textwidth]{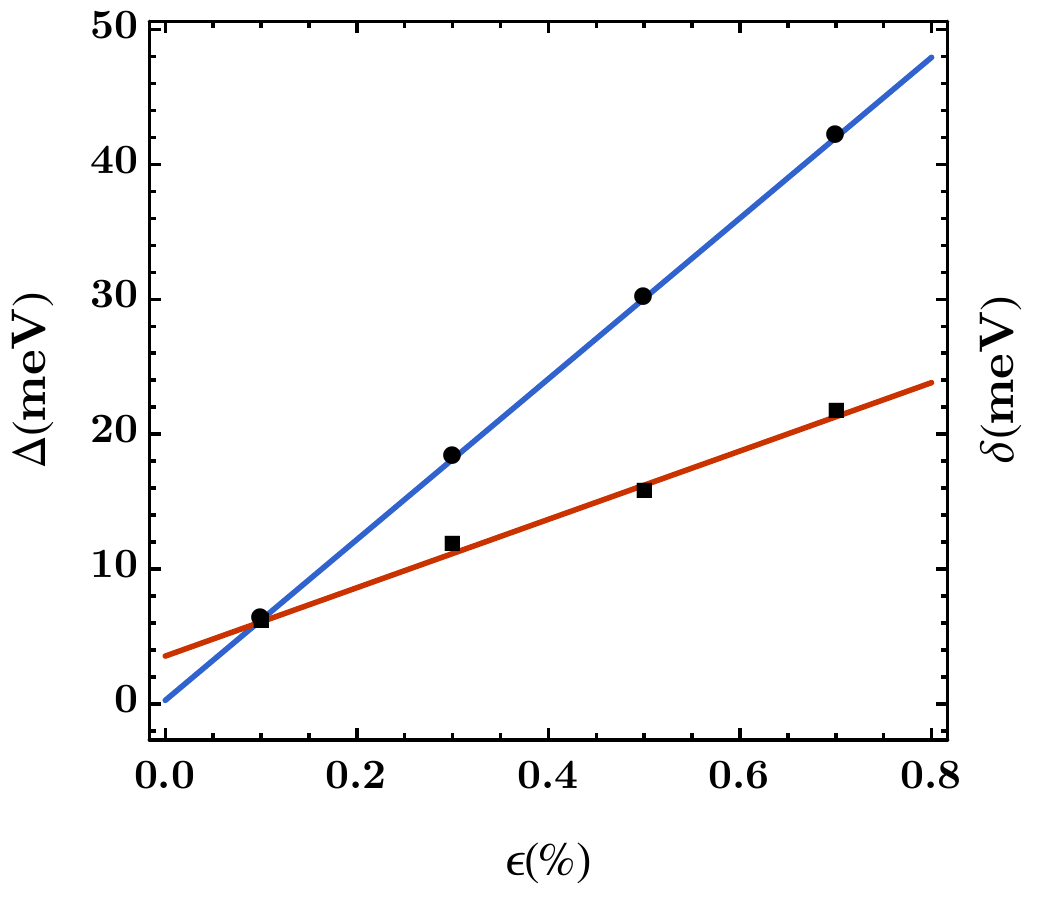}
\end{center}
\caption{The blue line shows the saturation values of the van Hove singularity splitting, labeled by $\Delta$, as a function of the magnitude of the uniaxial strain. The red line shows the maximal energy shift of the two Dirac crossings within a valley, $\delta$, as a function of strain magnitude. The fitting shows that both quantities are linearly proportional to the magnitude of uniaxial strain.}
\label{fig:fit}
\end{figure}

Secondly, we study the total bandwidth measured by the separation of the van Hove singularities in the conduction and valence bands, labeled by $\Delta$. Fig. \ref{fig:BW2} shows the evolution of $\Delta$ as a function of $\theta$ for several different $\epsilon$'s at fixed strain direction $\varphi=25^\circ$. An interesting feature is that the bandwidth is \textit{insensitive} to the twist angle for angles near the magic angle. We also observe that the saturation value of the bandwidth is linearly proportional to the magnitude of the strain as shown in Fig. \ref{fig:fit}, which suggests the strain effect is \textit{dominant} near the magic angle. In certain sense, the heterostrain provides an intrinsic kinetic energy cutoff, $\Lambda\sim \epsilon \hbar v_F$, for the system and stabilizes the bandwidth.


\begin{figure}
\begin{center}
\includegraphics[width=0.38\textwidth]{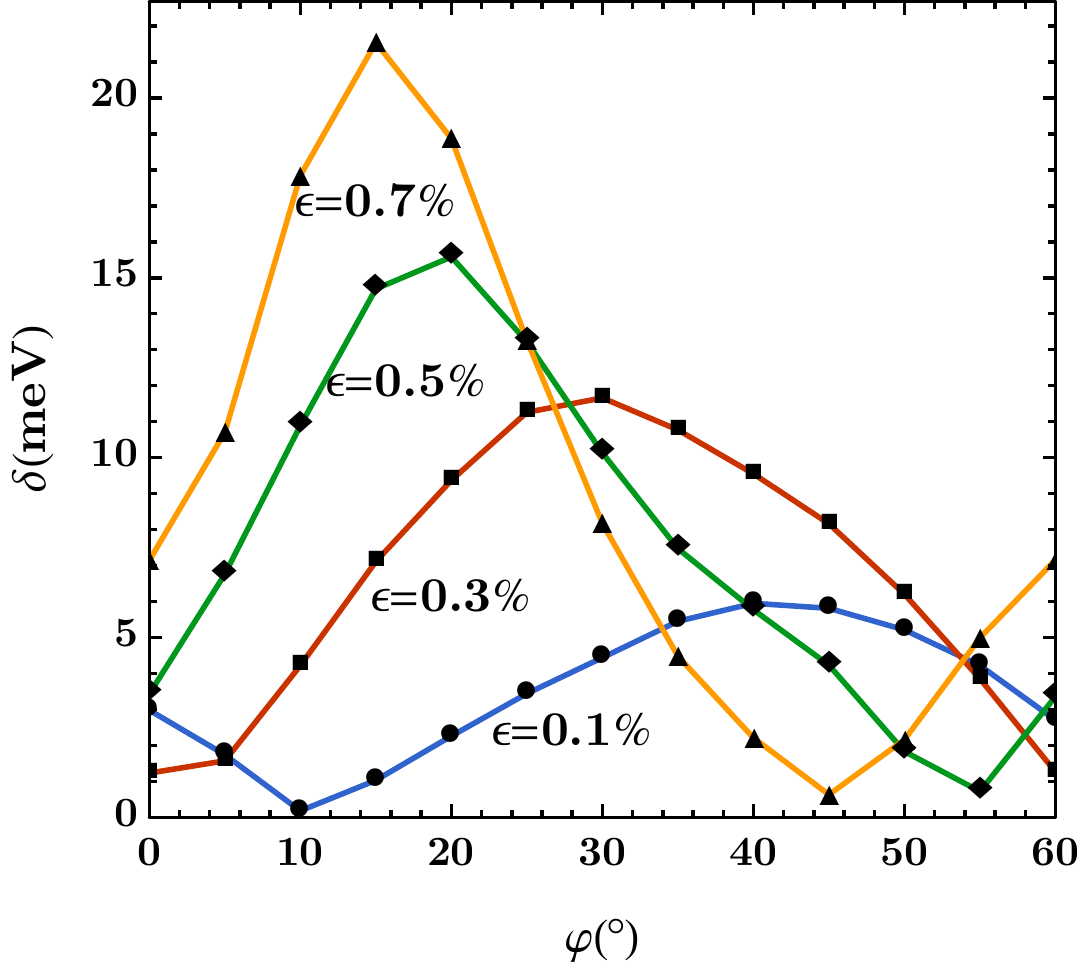}
\end{center}
\caption{The energy shift between two Dirac points in a valley, labeled by $\delta$, as a function of the direction of the strain. In this plot, we fix $\theta=1.05^\circ$ and choose 4 different strain magnitudes $\epsilon=0.1\%$(blue), $0.3\%$ (red), $0.5\%$ (green), $0.7\%$ (yellow). We notice that the energy shift of the Dirac fermions is very sensitive to the direction of the strain. The maximal value of the shift is linear proportional to the strain magnitude.}
\label{fig:shiftphi}
\end{figure}

Finally, we study the energy shift, labeled by $\delta$, of two Dirac points within a valley as a function of $\epsilon$ and $\varphi$ at fixed twist angle. As shown in Fig. \ref{fig:shiftphi}, the Dirac point shifts highly depend on the direction of the strain. We also find that the maximal value of the Dirac point shift is proportional to the magnitude of the strain. To our surprise, a quite small heterostrain such as $\epsilon=0.1\%$ can create a considerable splitting of the dirac fermions (on the order of $5$meV).

 Although a complete analytical understanding of heterostrain effects found in our calculation is currently lacking, we provide some intuition for the observed features in the band structures. In the unstrained twisted bilayer near the magic angle, the nearly flat energy dispersions arise from subtle interference effects between the mismatched energy dispersion within each layer  and electron hopping between layers. For example, the state at $\bold{K}_-$ from layer-1 would be coupled through the moir\'e interlayer tunneling to three states from layer-2 with the same energy. Similarly, the state at $\bold{K}_-$ from layer-2 is also coupled to three equal energy states in layer-1. We can crudely think of the magic angle is tuning the bilayer system to a critical situation where the kinetic energy scale vanishes. However, such ``interference condition" is explicitly violated by the heterostrain, which introduces a natural kinetic energy cutoff, $\Lambda\sim \epsilon \hbar v_F$, to the system. It is easy to see the energies of the three states from layer-2 that can couple to $\bold{K}_-$ state in layer-1 are now shifted precisely by $\Lambda\sim \epsilon\hbar v_F$ because of the heterostrain. This lifting of perfect interference is presumably the cause for the enlarged bandwidth and the Dirac fermion energy shift.

\subsection{Higher order van Hove singularities}

\begin{figure}
\begin{center}
\includegraphics[width=0.45\textwidth]{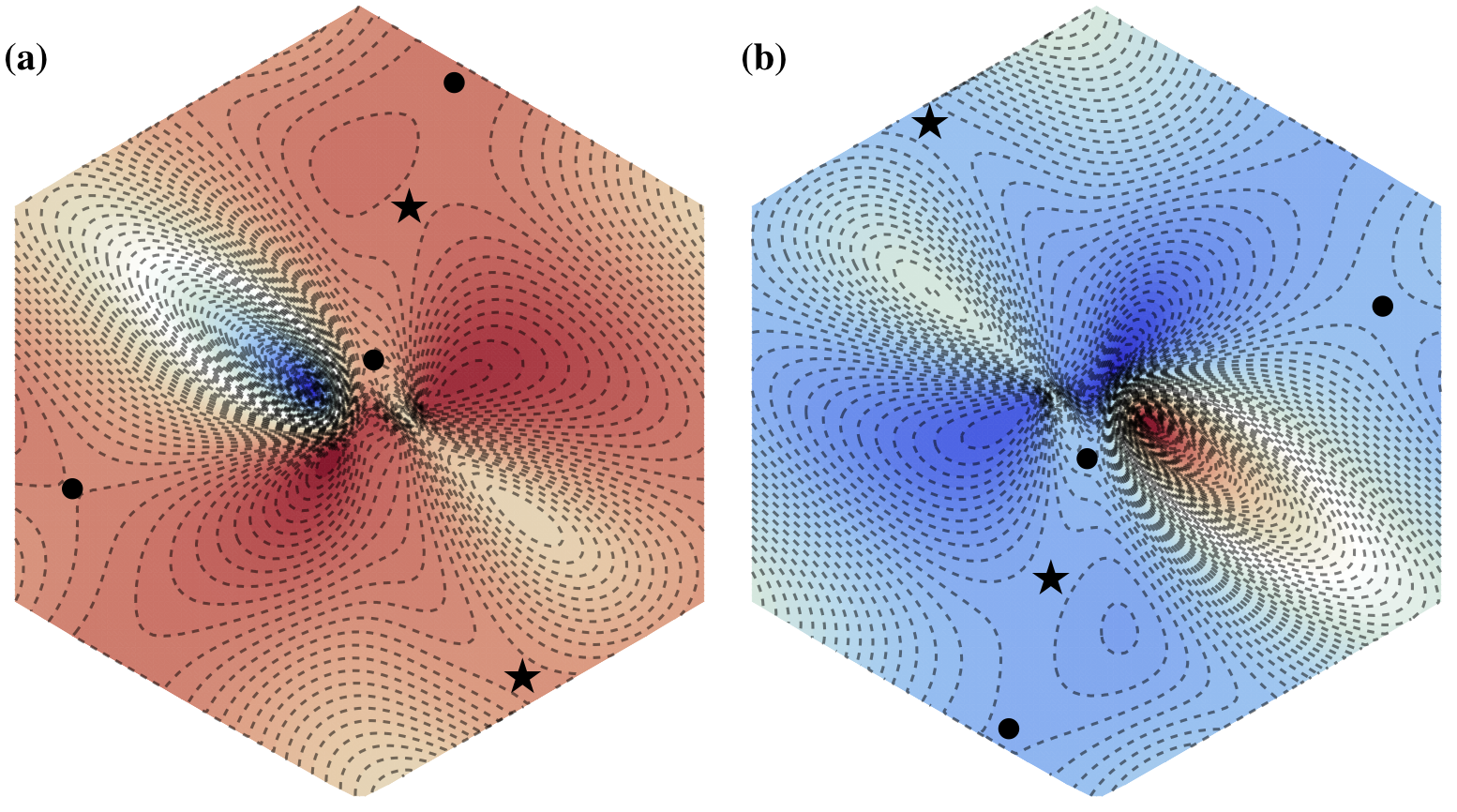}
\includegraphics[width=0.42\textwidth]{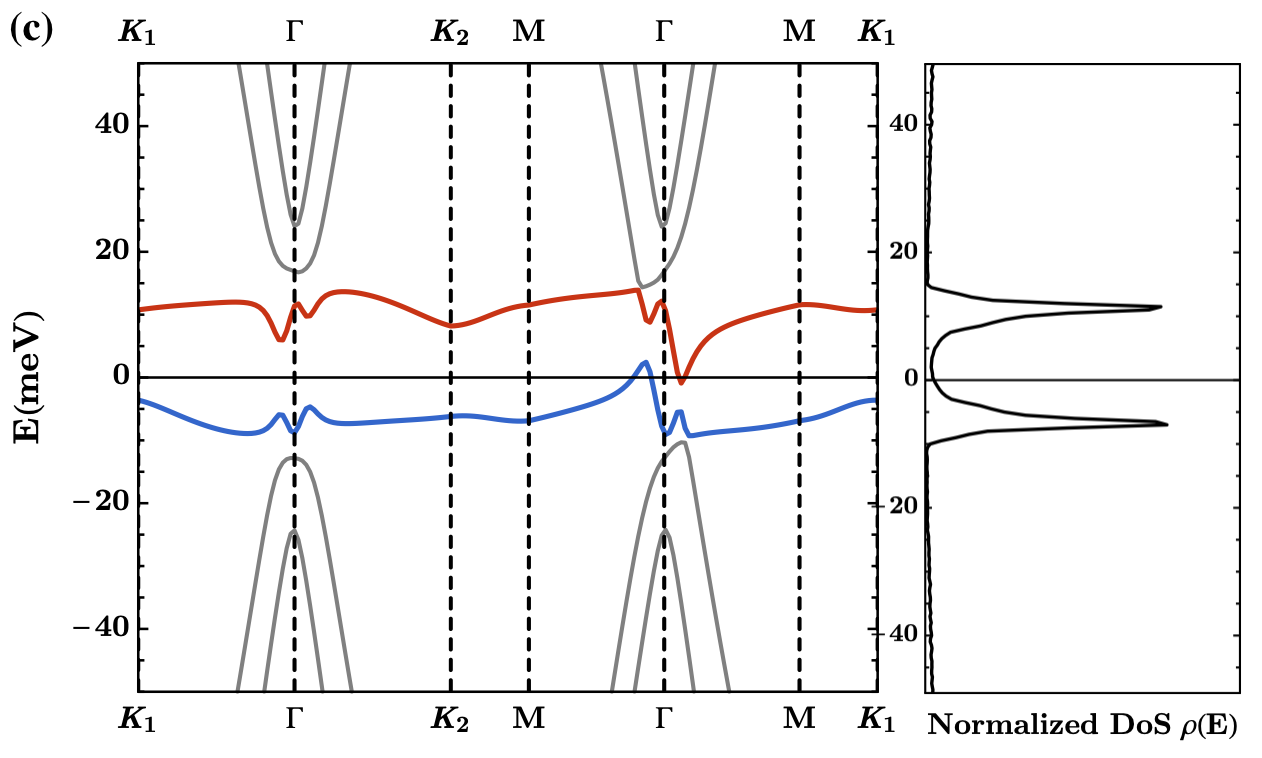}
\end{center}
\caption{(a)/(b) The equal energy contour plots for the valence/conduction bands with parameters $\theta=1.05^\circ$, $\epsilon=0.3\%$ and $\varphi=25^\circ$. We can identity multiple ordinary (dot) and higher order (star) van Hove singularities. Since these van Hove singularities happen to be nearby in energy, they contribute to a single enhanced density of state peak in each band as shown in (c). }
\label{fig:vHs1}
\end{figure}

\begin{figure}
\begin{center}
\includegraphics[width=0.45\textwidth]{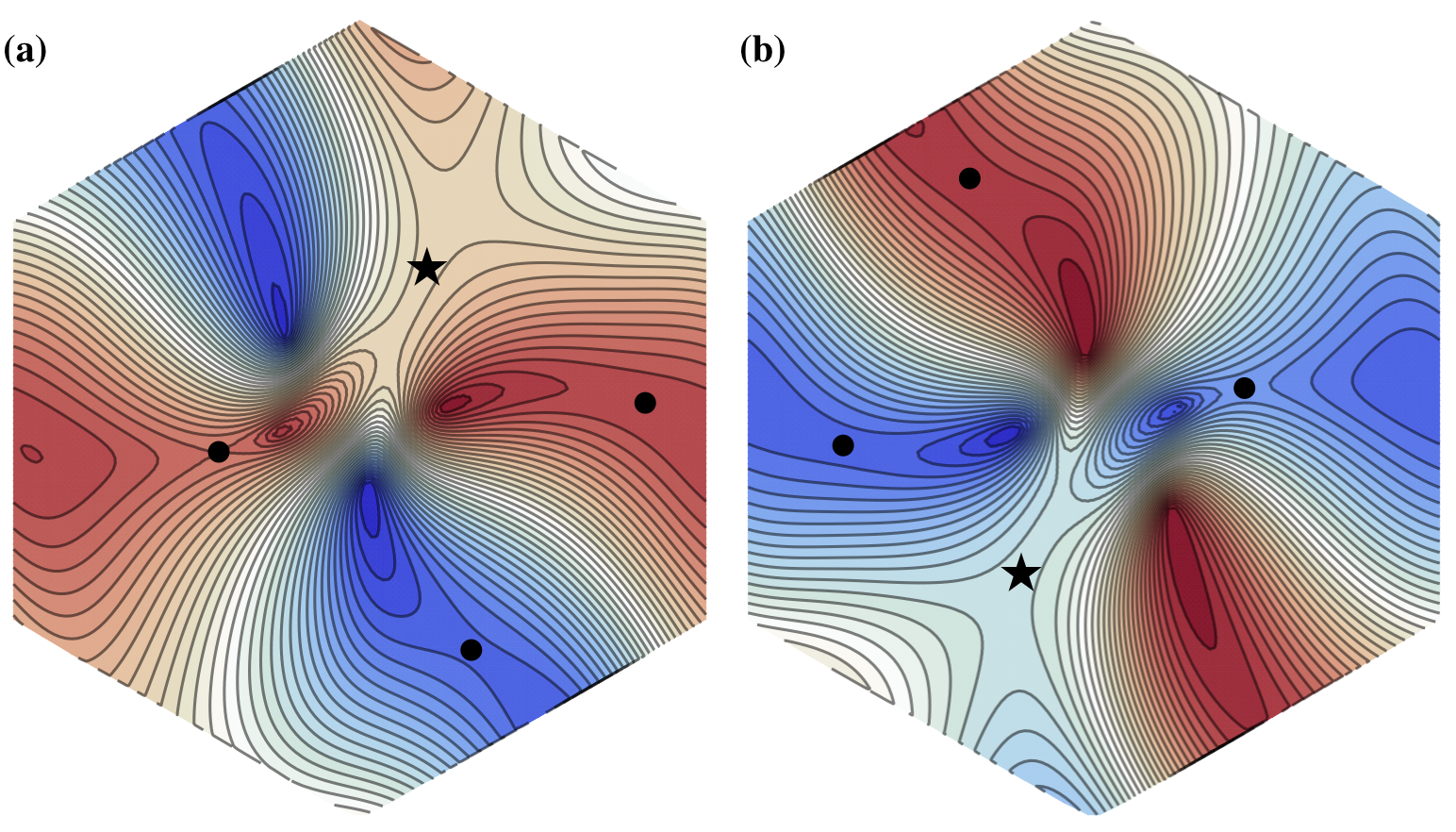}
\includegraphics[width=0.45\textwidth]{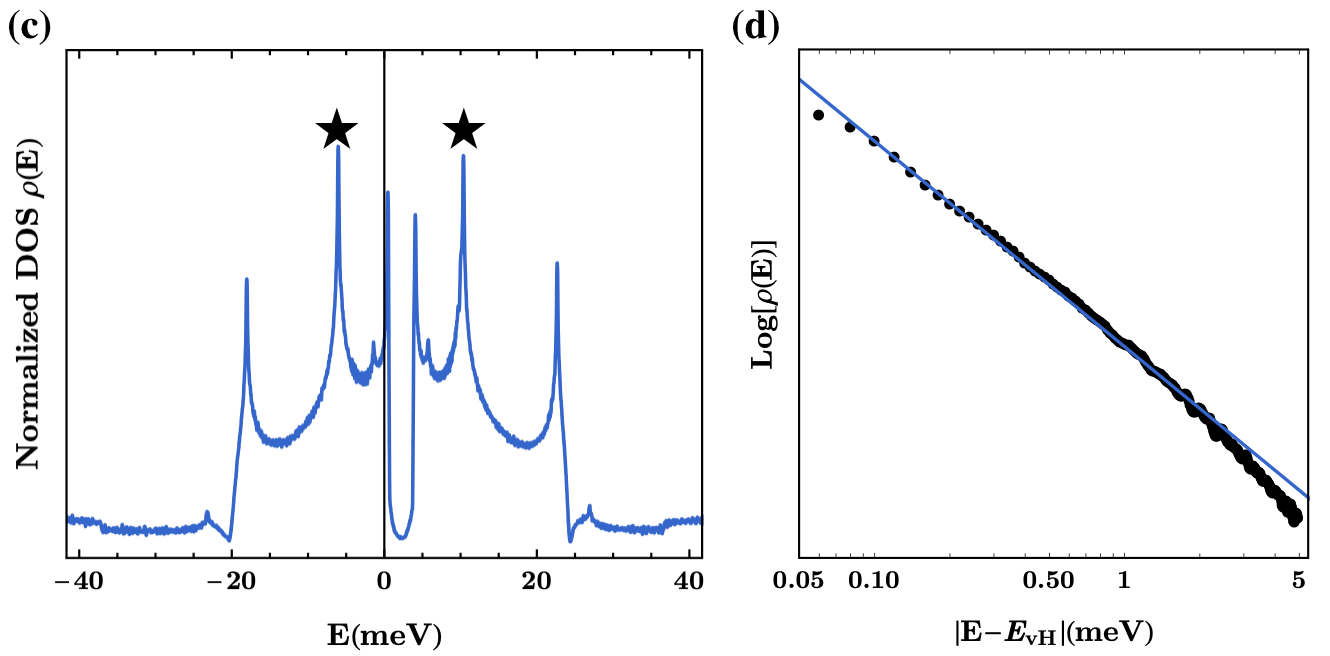}
\end{center}
\caption{(a)/(b) The equal energy contour plots for the valence/conduction bands with parameters $\theta=1.05^\circ$, $\epsilon=0.53\%$ and $\varphi=50^\circ$. We observe ordinary and higher order van Hove singularities in both valence and conduction bands. They contribute to the sharp peaks in the density of states in (c). In (d) we zoom in near the density of state peak at $E_{vH}\cong 10.4$meV and make a $log-log$ plot for $\rho(E)$ vs $|E-E_{vH}|$ for $E-E_{vH}>0$. The linear fitting indicates that the density of states has a power law divergence near the van Hove singularity, namely $\rho(E)\sim |E-E_{vH}|^{-\nu}$ with $\nu\cong 0.255$. This is the key feature for the higher order van Hove singularities.}
\label{fig:vHs2}
\end{figure}

As shown in Ref. \cite{FuHOVH}, with a single tuning parameter such as the twisted angle, one can achieve higher order van Hove singularities in bilayer graphene, which are perfect playgrounds for correlation-driven physics. Here we show that heterostrain is another effective way to generate higher order van Hove singularities. To demonstrate this, we show two examples in Fig. \ref{fig:vHs1} and \ref{fig:vHs2}. We find multiple ordinary and higher order van Hove singularities in both conduction and valence bands in these examples. The energies of the van Hove singularities are not required to be the same in general because of the low symmetry of the heterostrained system. Therefore, we may observe multiple peaks in the density of states for certain conditions as shown in Fig. \ref{fig:vHs2}.

\section{Heterostrain in bilayer transition metal dichalcogenide}

\subsection{Heterostrain induced flat bands}

For monolayer TMD, the model hamiltonian for one valley can be written as the following\cite{XDTMD1,YWTMD1,Tstrain}
\beq
h_l(\bold{k})=\sum_{\xi=\pm}-\hbar v_F/a [(\mathbb{I}+\mathcal{E}_l^T)(\bold{k}-\bold{D}_{l,\xi})]\cdot(\xi\sigma^x,\sigma^y)+\frac{m}{2}(\sigma^z+1),
\eeq
where $\sigma$ matrices act on sublattice space and there is no spin degeneracy because of the large spin-orbital coupling. This hamiltonian generally applies for a large class of TMD materials\cite{XDTMD1,YWTMD1, Tstrain}. In this paper, we take WSe$_2$ as a representative for TMD materials. For WSe$_2$, it is estimated that $\hbar v_F/a\cong 1.1$eV and $m\cong 1.2$eV. Similar as the graphene case, with strains, the Dirac points are shifted and their locations in momentum space are again given by Eq. \ref{eq:dirac} and \ref{eq:shift}. It is found from first principle that the drift parameter $\beta$ is around $2.3$ for WSe$_2$\cite{TMDstrain, drifts}. Of course, it is hard to pin down the exact value of the $\beta$ and it also varies for different TMD materials.  Therefore, we treat $\beta$ as a potential tuning parameter to extract general features of the moir\'e band structure.

In this section, we consider bilayer transition metal dichalcogenide system with only heterostrains for simplicity. It turns out that the bilayer TMD system is a desirable platform for engineering flat bands with strains. We restrict ourselves to a class of heterostrain which is called volume preserving strain. The general form is given by,
\begin{eqnarray}
S_v=
\epsilon\begin{pmatrix}
\cos\varphi & \sin\varphi\\
\sin\varphi & -\cos\varphi
\end{pmatrix}.
\label{eq:polar}
\end{eqnarray}
Physically, $S_v$ describes that the material is strained by $\epsilon$ along $\varphi$ direction and by $-\epsilon$ along the normal direction. We consider the bilayer system starting from \textit{AA stacking} configuration, where the inversion symmetry is not present. The \textit{AB stacking} configuration is an interesting case which we will address in a separate paper. Applying a generic $S_v$ explicitly breaks all the point group symmetries of the system.

For bilayer WSe$_2$ moir\'e superlattices starting from \textit{AA stacking} configuration, we need to include both the interlayer tunneling and the intralayer potential in Eq. \ref{eq:H}. The interlayer tunneling $T(\bold{r})$ has the same form as Eq. \ref{eq:tunneling}. It is estimated in Ref.\cite{MacDTMD1} that, for bilayer WSe$_2$, the tunneling parameters are $u\cong1.1$meV, $u''\cong9.7$meV and $u'$ approximately zero because of the large band gap.   The intralayer potential has the following form\cite{MacDTMD1}
\beq
V_l(\bold{r})=\sum_{i=1,2,3}\left(\begin{matrix}
V_c e^{i(\bold{g}_i\cdot\bold{r}+(-1)^l\phi_c)}& 0 \\
0 & V_ve^{i(\bold{g}_i\cdot\bold{r}+(-1)^l\phi_v)}
\end{matrix}\right)+h.c.,
\label{eq:potential}
\eeq
where $\bold{g}_3=-(\bold{g}_1+\bold{g}_2)$. For WSe$_2$, the parameters are $V_c\cong 6.8$meV, $V_v\cong 8.9$meV, $\phi_c\cong 89.7^\circ$ and $\phi_v\cong 91^\circ$. These data vary slightly for other TMD materials. Thus, we will focus on generic features of the electron band structure which do not rely much on the precise values of these parameters.
\begin{figure}
  \begin{center}
  \includegraphics[width=0.48\linewidth]{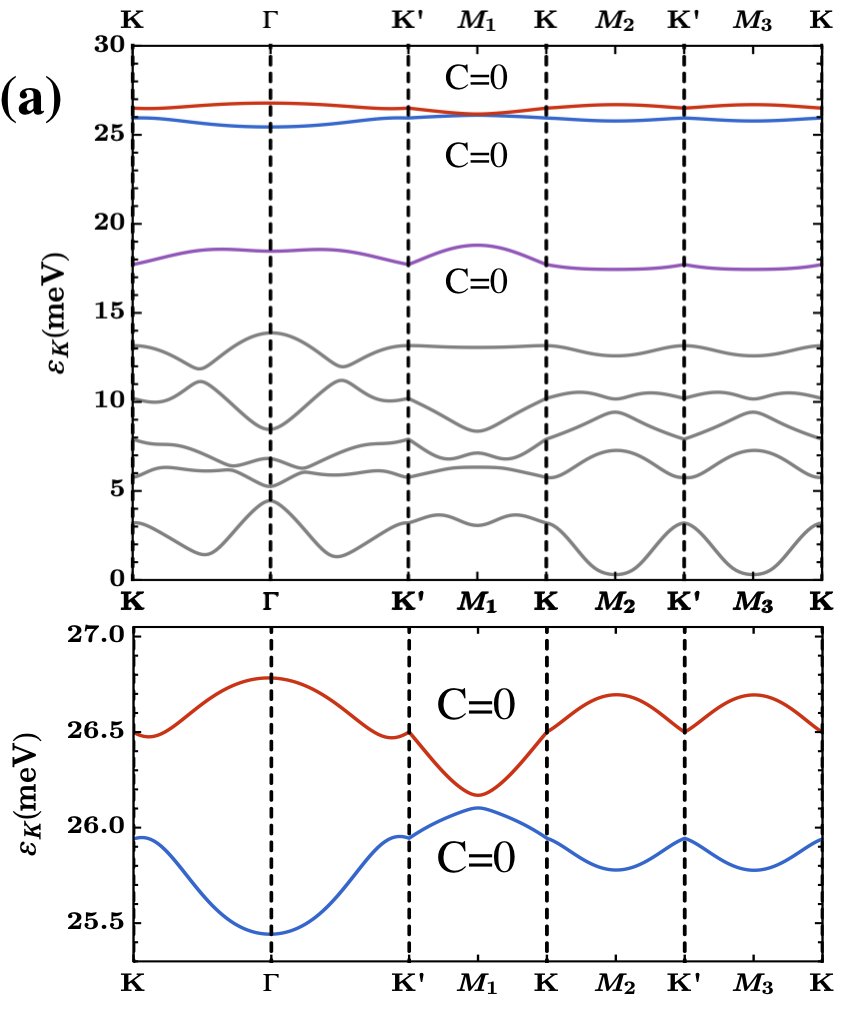}
  \includegraphics[width=0.48\linewidth]{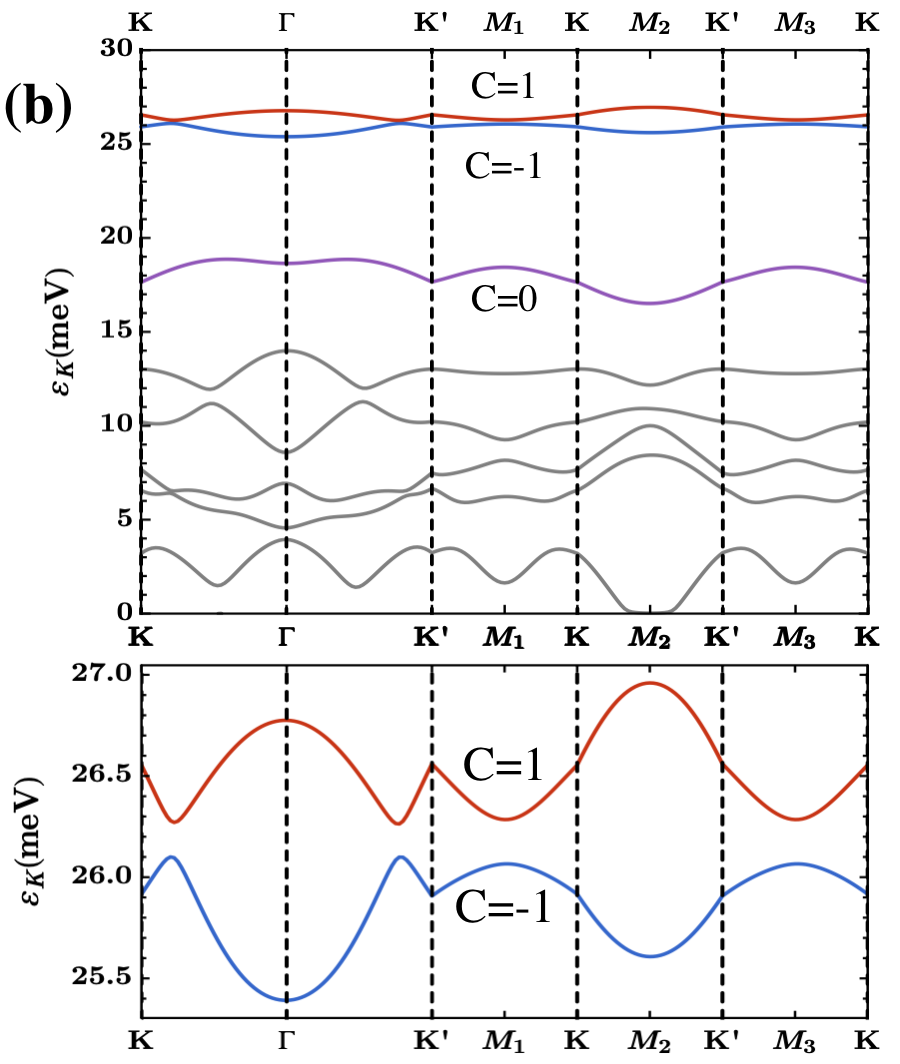}
    \end{center}
\caption{We focus on the band structures for $\bold{K_-}$ valley of bilayer WSe$_2$ near the top of the valence band. (a)/(b) show band structures with volume preserving heterostrain at $\epsilon=1.5\%$ and $\varphi=0^\circ / 30^\circ$ respectively. (We take $\beta_{TMD}\cong 2.3$.) We notice that the first three bands are quite flat. The top two bands are close in energy and well separated from other bands with gap $\Delta\sim 8$meV. We observe that for $\varphi=0$, the first two valence bands are topologically trivial, while for $\varphi=30^\circ$ they carry $\pm 1$ chern number.}
\label{fig:TMD-band}
\end{figure}

We show the band structure for bilayer WSe$_2$ with heterostrain calculated from the continuum model in Fig. \ref{fig:TMD-band}. We find generic flat bands near the top of the valance band for small strain magnitude. The top two valence bands are close in energy and have very small bandwidth compared to the large gap separating them to the rest of the spectrum. The third band is also quite flat and energetically separated from other bands. Increasing strain can enlarge the bandwidth and band gap.

\begin{figure}
  \begin{center}
   \includegraphics[width=0.47\linewidth]{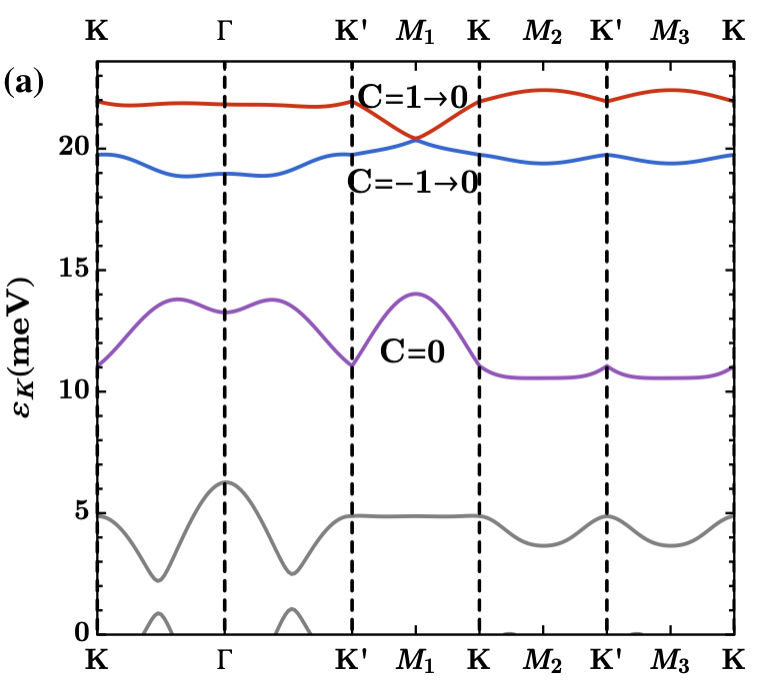}
   \includegraphics[width=0.47\linewidth]{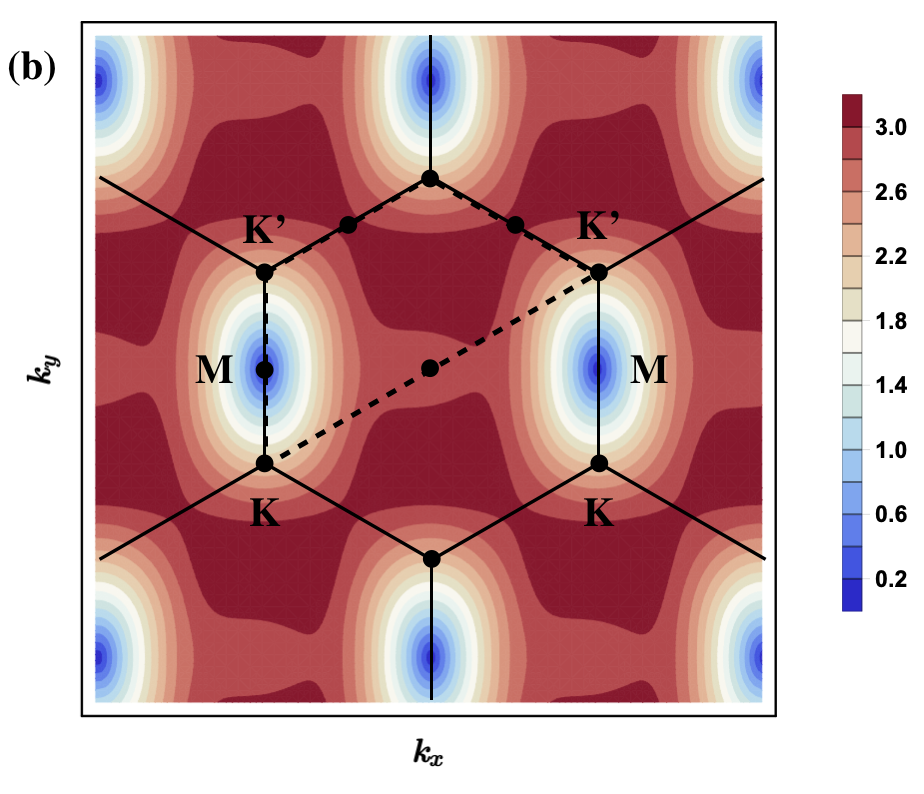}
   \includegraphics[width=0.46\linewidth]{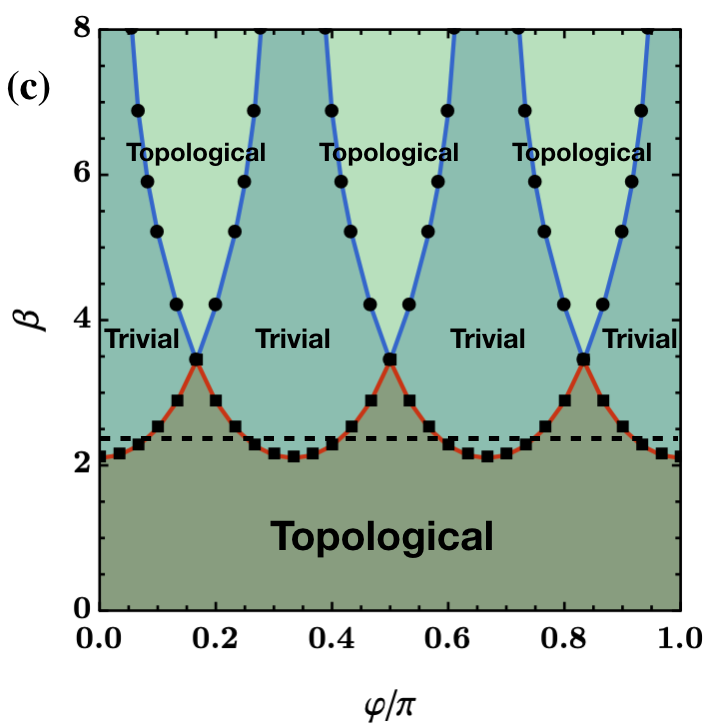}
   \includegraphics[width=0.47\linewidth]{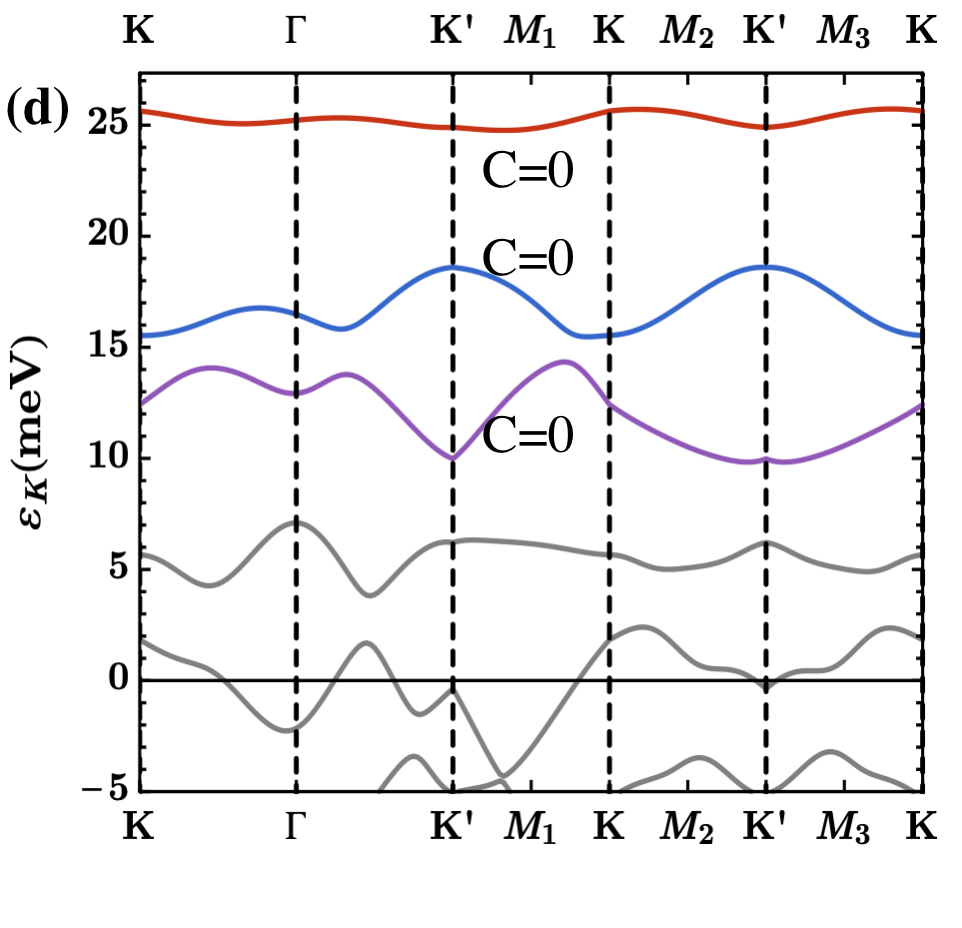}
    \end{center}
\caption{(a) shows the band structure of $\bold{K_-}$ valley with parameters $\beta=2.10$, $\epsilon=2\%$ and $\varphi=0$. The system is at a topological phase transition. A Dirac crossing is found at one of the $\bold{M}$ points in the Brillouin zone. The three $\bold{M}$ points are not equivalent because of the lack of $C_3$ rotation symmetry. (b) shows the contour plot of the gap between the first two bands with the same parameters. (c) Phase diagram of the top two valence bands as a function of $\beta$ and $\varphi$. There is an interesting pattern of topological and trivial states (topological state refers to $\pm1$ chern number for the top two bands and trivial state refers to $0$ chern number) as we tune the angle $\varphi$ from $0$ to $\pi$. $\varphi=0$ and $\pi$ are physically equivalent by an exchange of the two layers. The dash line is the estimated value of $\beta\cong 2.30$ for WSe$_2$. The phase diagram is obtained with $\epsilon=2\%$. However, the phase diagram is \textit{insensitive} to the strain magnitude for small strain $<3\%$. (d) The band structure for $\epsilon=2\%$ and $\varphi=0$ with an interlayer bias $\Delta V=10$meV. The displacement field separates a nearly flat band on the top of the spectrum. }
 \label{fig:betaphi}
\end{figure}

The strain angle $\varphi$ provides an additional nob for engineering the band structure. Now we focus on the top two valence bands. In Fig. \ref{fig:betaphi} (c), we show the phase digram as a function of $\beta$ and $\varphi$ (the phase diagram is not sensitive to the magnitude of the strain $\epsilon$ for small $\epsilon$). Interestingly, we find an alternating pattern of topological and trivial bands as a function of $\varphi$. The threefold periodicity of the phase diagram in $ \varphi $ is due to the emergent threefold rotational symmetry of interlayer coupling (\ref{eq:tunneling}) and intralayer potential (\ref{eq:potential}). We show the band structure of a critical point at $\beta_c\cong 2.10$ and $\varphi=0$. The topological phase transition closes the gap with a Dirac dispersion at one of the $M$ points in the Brillouin zone.

It is worth to mention the displacement field is a good way to control the relative position of the first three bands. With considerable interlayer bias, we can make one band well separated from other bands on the top of the spectrum. In addition, this band is topologically trivial. An example of such band structure is shown in Fig. \ref{fig:betaphi} (d). As expected, the wavefunction of the top band is mostly polarized in one layer.

\subsection{Pressure tunned topological transition}

\begin{figure}
  \begin{center}
  \includegraphics[width=0.48\linewidth]{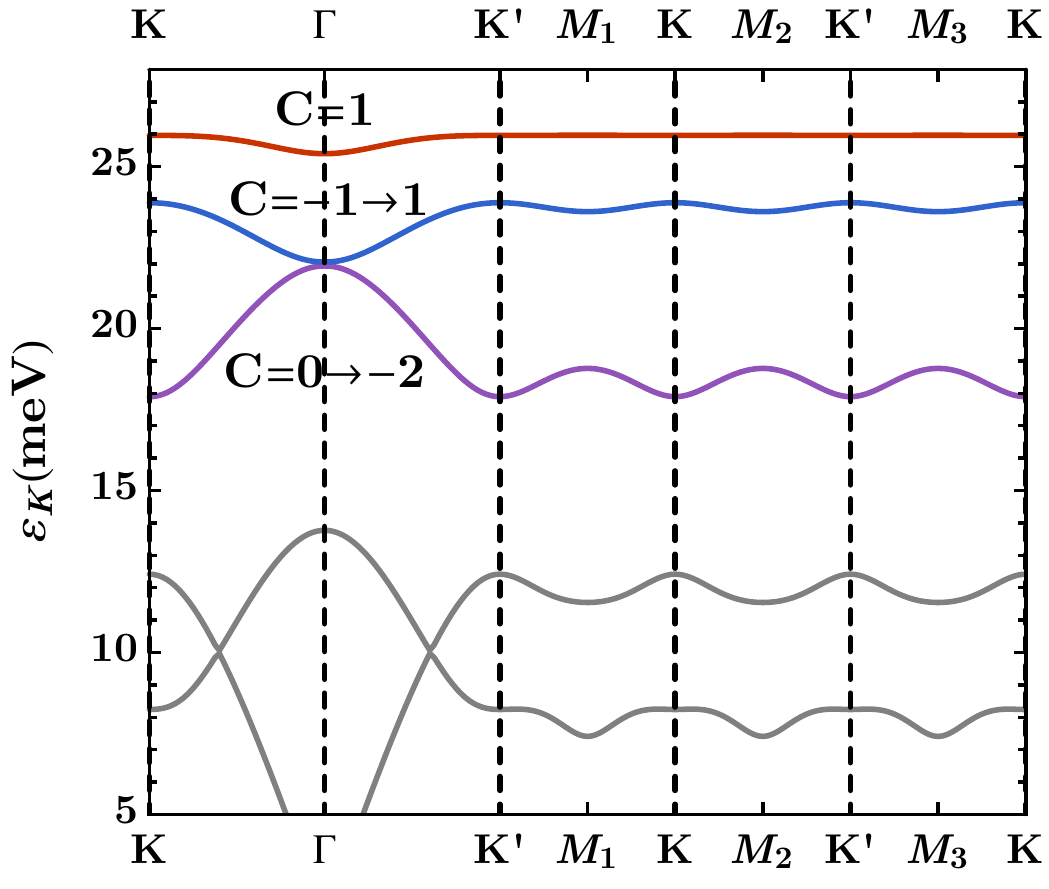}
  \includegraphics[width=0.48\linewidth]{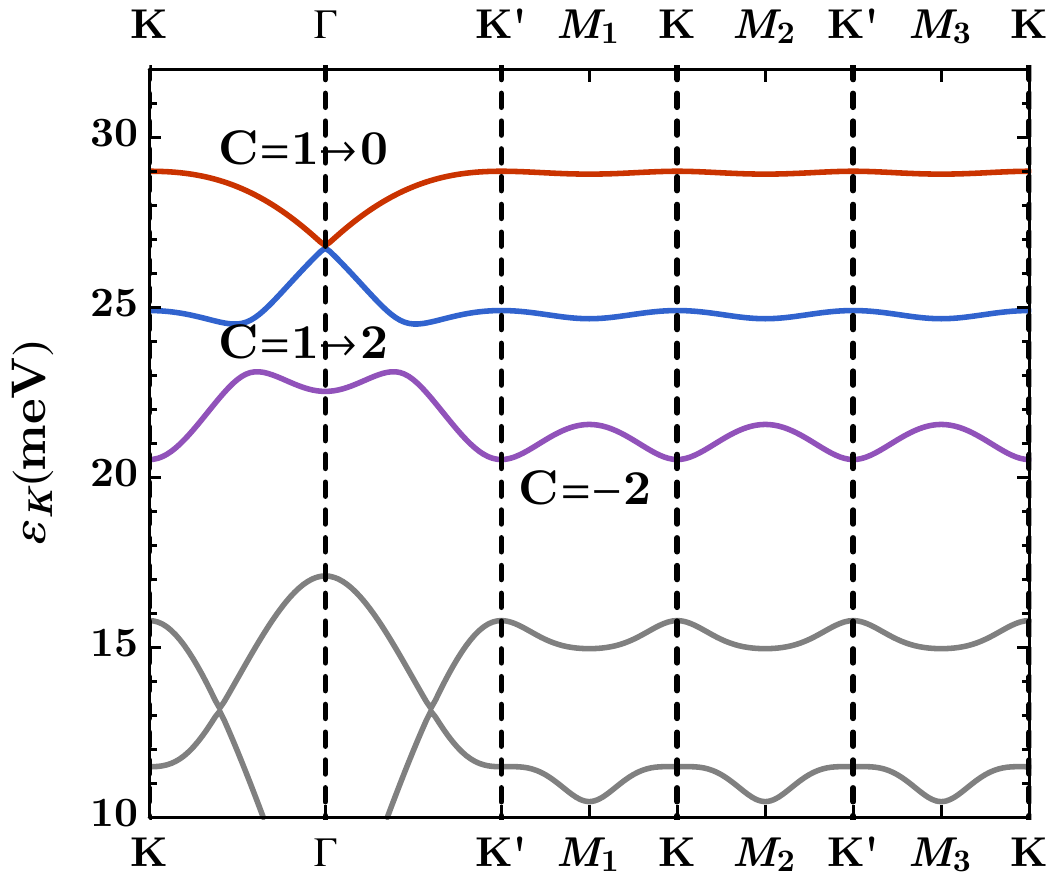}
    \end{center}
\caption{We consider twisted bilayer WSe$_2$ at $\theta=1^\circ$ with pressure. As the parameter $p$ is tuned from $1$ to $1.5$, there are two successive topological phase transitions. (a) At $p\cong 1.195$, the second band and third band touches with quadratic dispersion at $\Gamma$ point, which changes the Chern number by 2. (b) At $p\cong1.410$, the first and second band close gap with a Dirac dispersion, which changes the Chern number by 1.}
 \label{fig:TPT}
\end{figure}

Pressure on the bilayer systems changes the interlayer distance and in return adjusts the relative ratio between the strength of the interlayer tunneling and intralayer potential energy, which is another way to design the moir\'e band structure. To demonstrate this effect, we consider twisted or heterostrained bilayer TMD with pressure. Pressure can be modeled by introducing a phenomenological parameter $p$ in the model Eq. \ref{eq:H}, $T(\bold{r})\rightarrow pT(\bold{r})$, which represents the relative ratio of interlayer tunneling and intralayer potential. For twisted bilayer TMD at twist angle $\theta\cong 1^\circ$, we find multiple topological phase transitions between the top three moir\'e bands as we increase $p$ from $1$ to $1.5$ as shown in Fig. \ref{fig:TPT}. At $p\cong 1.195$, there is a quadratic band touching at $\Gamma$ in moir\'e Brillouin zone between the second and third band which shifts the Chern numbers of the two bands by $\pm 2$. At $p\cong 1.41$, the top two bands touch at $\Gamma$ with a Dirac dispersion. The top band becomes topologically trivial for $p> 1.41$. In this system, we can get topological flat bands with chern number $\pm 2$. Similar effects also appear in heterostrained cases.

\subsection{Effective model for the flat bands}

With small heterostrains, for instance $\epsilon\cong 1.5\%$ in Fig. \ref{fig:TMD-band}, for each valley the two top valence bands which are close in energy and very flat (total bandwidth $W\cong 1.4$meV), and they are well separated from the rest of the bands (gap $\Delta\cong 8$meV). One can estimate the Coulomb interaction scale on the moir\'e superlattice $V= e^2/(4\pi \epsilon a_M)\cong 4.2$meV, with static dielectric constant of WSe$_2$ $\epsilon\cong 15.3$\cite{dielectric} and moir\'e superlattice constant $a_M\cong 67 a\cong 23.4$nm for $\epsilon_{xy}=1.5\%$. The parameters of the system are in the limit $\Delta\gg V\gg W$. In addition, the total Chern number for the top two valence bands is trivial. Therefore, we can propose an effective model just for the top two bands. Taking into account of the valley/spin degeneracy, the simplest guess is a 2-band Hubbard model on a triangular lattice, which can be schematically written as the following
\bea
\nn
H_{eff}=&\sum_{\alpha=1}^{2}\sum_{i,j}\sum_{v=\pm}t^\alpha_{ij}c^\dagger_{i,\alpha,v} c_{j,\alpha,v} \\
&+\sum_i \sum_{\alpha=1}^2U_\alpha n_{i,\alpha}(n_{i,\alpha}-1)+...,
\eea
where $n_{i,\alpha}=\sum_{v=\pm}c_{i,\alpha,v}^\dagger c_{i,\alpha,v}$. The two valleys are found to be degenerate in energy, therefore, have an approximate $SU(2)$ symmetry in the small intervalley-scattering limit. The hopping terms are spatially anisotropic depending on the heterostrain. Half filling the two bands may lead to interesting correlated states, such as valley/spin polarized state, which spontaneously breaks time reversal symmetry.

Aside from the top two valence bands, we can also focus on the third band which is also quite flat and energetically separated from other bands. The Chern number is again trivial for this band with current parameters. The effective model for this band will be a single band triangular lattice $SU(2)$ Hubbard model with anisotropic hopping terms. Similar model could also describe the physics of the displacement field separated flat band in Fig. \ref{fig:betaphi} (d). This model is previously suggested to be the effective model of the quasi-two-dimensional
organic charge transfer salts, which potentially host exotic spin liquid states\cite{Organic1,Organic2}.

Finally, we emphasis that the Coulomb interaction scales with the strain as $V\sim 1/a_M\sim \epsilon$ while the bandwidth of the moir\'e bands scales with the strain as $W\sim (1/a_M)^2 \sim \epsilon^2$. Therefore, different strains can produce Hubbard models with different $t/U$ ratio.

\section{Discussion}

In this paper, we explore the effects of uniaxial heterostrain in twisted bilayer graphene. It is clear that uniaxial heterostrain generically broadens the bandwidth of the nearly flat bands at magic angle, which could be the reason for the observed large energy separation of van Hove singularities in the STM experiment\cite{TBGstm}. Large bandwidth may not be desirable for correlation physics. Therefore, future experiments should try to minimize the heterostrain between layers. On the other hand, heterostrain could be helpful to generate higher order van Hove singularities, which is a good platform to enhance correlation effects. A systematic study of the evolution of the van Hove singularities is an interesting but involved subject. Another natural question to ask is whether there are situations in which heterostrain helps to flatten the band dispersions rather than to broaden the bands. To answer these questions, we need more comprehensive studies where optimization methods such as machine learning could be helpful. On the other hand, a full analytical understanding of the strain effect is also demanded for future study.

The other effects of heterostrain are shifting the energies of the two Dirac crossings within one valley and greatly enhancing the Dirac velocity. This could potentially explain the observed 4-fold Landau level degeneracy in magneto transport experiments\cite{TBGexp1,TBGexp2,TBGexp3}.  One natural consequence of such scenario is that there will be finite electron/hole fermi surfaces at the charge neutrality point. Nonetheless, due to the enhanced Dirac velocity, the fermi surfaces are estimated within our continuum model to be really small and could be difficult to resolve with current experimental precision. To get a more accurate estimation theoretically, one needs to include various subtle effects such as lattice relaxations, which is beyond the scope of the current paper but a good subject for future investigations.

We also show heterostrain is a good way of creating and tuning flat bands in bilayer TMD systems. There is no requirement of a ``magic" strain here. The band structure is easily controlled by the strain direction, pressure and displacement field.

\section*{Acknowledgement} We thank Abhay Pasupathy, Cory Dean,  Matt Yankowitz and Brian LeRoy for stimulating discussions and communications. ZB is supported through Pappalardo fellowship at MIT. NY and LF are supported by DOE Office of Basic Energy Sciences, Division of Materials Sciences and Engineering under Award
DE-SC0018945. LF is partly supported by the David and Lucile Packard Foundation.

\bibliography{strain}

\end{document}